\documentclass{emulateapj}

\begin{document}

\shortauthors{Luhman \& Sheppard}
\shorttitle{High Proper Motion Objects from WISE}

\title{Characterization of High Proper Motion Objects from
the {\it Wide-field Infrared Survey Explorer}\altaffilmark{1}}

\author{
K. L. Luhman\altaffilmark{2,3} and
Scott S. Sheppard\altaffilmark{4}}

\altaffiltext{1}
{Based on data from the {\it Wide-field Infrared Survey Explorer},
the Two Micron All-Sky Survey, the NASA Infrared Telescope Facility,
Gemini Observatory, the SOAR Telescope, and the Magellan Telescopes.}

\altaffiltext{2}{Department of Astronomy and Astrophysics,
The Pennsylvania State University, University Park, PA 16802, USA;
kluhman@astro.psu.edu}
\altaffiltext{3}{Center for Exoplanets and Habitable Worlds, The 
Pennsylvania State University, University Park, PA 16802, USA}

\altaffiltext{4}{Department of Terrestrial Magnetism, Carnegie Institution
of Washington, 5241 Broad Branch Rd. NW, Washington, DC 20015, USA}

\begin{abstract}

We present an analysis of high proper motion objects that we have found
in a recent study and in this work with multi-epoch astrometry from the
{\it Wide-field Infrared Survey Explorer (WISE)}. 
Using photometry and proper motions from 2MASS and {\it WISE}, we have
identified the members of this sample that are
likely to be late type, nearby, or metal poor. We have performed optical and
near-infrared spectroscopy on 41 objects, from which we measure spectral
types that range from M4--T2.5. This sample includes 11 blue L dwarfs
and five subdwarfs; the latter were also classified as such in the
recent study by Kirkpatrick and coworkers.  Based on their spectral
types and photometry, several of our spectroscopic targets may have
distances of $<20$~pc with the closest at $\sim12$~pc.
The tangential velocities implied by the spectrophotometric distances
and proper motions indicate that four of the five subdwarfs are probably
members of the Galactic halo while several other objects, including
the early-T dwarf WISE J210529.08$-$623558.7, may belong to the thick disk.

\end{abstract}

\keywords{
brown dwarfs ---
infrared: stars ---
proper motions --- 
solar neighborhood ---
stars: low-mass
}

\section{Introduction}

Proper motions have long been used to identify members of the solar
neighborhood.
Measurements at optical wavelengths have been sensitive to stellar masses
\citep{bar16,wol19,ros26,van44,gic71,luy79,sal03,ham04,lep02,lep03,lep05a,lep08a,sub05,boy11,fin12}
while data in near-infrared (IR) bands have reached substellar objects
\citep{art06,art10,dea05,dea09,dea11,dea07,sch09e,sch10d,she09,kir10,smi14}.
The near-IR proper motion surveys have been enabled primarily by 
the Two Micron All-Sky 
Survey \citep[2MASS,][]{skr06}, the Deep Near-Infrared Survey of the Southern
Sky \citep[DENIS,][]{ep99}, the Sloan Digital Sky Survey \cite[SDSS,][]{yor00},
the United Kingdom Infrared Telescope Infrared Deep Sky Survey
\citep[UKIDSS,][]{law07}, and Pan-STARRS1 \citep{kai02}.

In 2010, the imaging available for proper motion surveys was extended to mid-IR
wavelengths by the {\it Wide-field Infrared Survey Explorer}
\citep[{\it WISE},][]{wri10}.
Proper motions have been measured by combining the {\it WISE} astrometry with
optical and near-IR catalogs
\citep{liu11,sch11,sch12,giz11a,giz11b,giz12,cas12,cas13,bih13}
and by employing only the multiple epochs of data obtained by {\it WISE}
\citep{luh13,luh14a,tho13,wri14,kir14}.
In one of the latter surveys, \citet{luh14a} searched for a distant
companion to the Sun via the large parallactic motion that it would exhibit.
During the course of that study, several hundred new high proper motion objects
were found. In this paper, we investigate the nature of those objects by using
photometry and proper motions from 2MASS and {\it WISE} to 
identify the ones that are likely to be late type, nearby, or metal poor
(Section~\ref{sec:phot}). We then use spectroscopy (Section~\ref{sec:spec})
and kinematics (Section~\ref{sec:kin}) to characterize in more detail a subset
of that sample, focusing on the most promising candidates for L/T dwarfs
and subdwarfs.

\section{Photometric Analysis}
\label{sec:phot}

\subsection{Sample of High Proper Motion Objects}

\citet{luh14a} used multi-epoch astrometry from {\it WISE} to identify 762
high proper motion objects\footnote{The sample from \citet{luh14a} was
intended to contain only new high proper motion objects. However,
\citet{luh14a} overlooked the proper motion survey by \citet{pok04},
who had already identified 102 of the 762 objects presented by \citet{luh14a}.}.
For this study, we consider the 761 sources
that were detected by 2MASS because our analysis relies on both 2MASS
and {\it WISE} photometry. The one remaining object that is absent from
2MASS, WISE J085510.83$-$071442.5, has been characterized by \citet{luh14b}.
We also examine an additional 20 high proper motion objects that are presented
in Table~\ref{tab:hipm}, one of which (WISE J175510.28+180320.2) has
already been reported by \citet{gri12} and \citet{mac13a}.
These objects did not satisfy the criterion of $\mu$/$\sigma_\mu>5$ 
for inclusion in the sample from \citet{luh14a}, but they did 
exceed that threshold in an early version of the survey by \citet{luh14a}
that used the preliminary release of data for the Post-Cryo phase, and at
that stage they were confirmed as high proper motion stars using detections 
from 2MASS.
Although all of these 20 sources appear in 2MASS images, some of them lack
entries in the 2MASS Point Source Catalog, as noted in Table~\ref{tab:hipm}.

%The sample from \citet{luh14a} was intended to contain only new high proper
%motion objects. However, \citet{luh14a} overlooked the previous
%proper motion survey by \citet{pok04}, who had already identified 102
%of the 762 objects presented by \citet{luh14a}.

%The sample from \citet{luh14a} was intended to contain only new high proper
%motion objects. However, while compiling previously known proper motion stars
%for exclusion from the sample, \citet{luh14a} overlooked the catalog from
%\citet{pok04},  The latter had already identified 102 of the 762 objects 
%from \citet{luh14a}.

%The sample from \citet{luh14a} was intended to contain only new high proper
%motion objects. However, 102 of the 762 objects from \citet{luh14a} were
%previously identified by \citet{pok04}.

\citet{kir14} have also performed a search for high
proper motion objects with data from {\it WISE}, arriving at a sample of 3525
sources with motions confirmed by detections in other surveys like 2MASS.
They described the overlap between their sample and the one from
\citet{luh14a} and compared the distributions of magnitudes and proper motions
between the two samples.
We make one additional comment regarding that comparison.
As illustrated in Figure~22 from \citet{kir14}, most of the stars that are
in \citet{kir14} but not in \citet{luh14a} have motions of
$\lesssim0\farcs3$~yr$^{-1}$.
This difference is a reflection of the fact that \citet{luh14a} excluded
{\it WISE} sources with 2MASS counterparts within $3\arcsec$, which
corresponds to $\mu\lesssim0\farcs3$~yr$^{-1}$, in order to focus on objects
with larger motions, particularly those moving fast enough to be companions
of the Sun ($\gtrsim4\arcsec$~yr$^{-1}$).

\subsection{Common Proper Motion Companions}
\label{sec:cpm}

In \citet{luh14a}, the 2MASS counterpart of WISE~J163348.95$-$680851.6 was
identified as 2MASS~J16334908$-$6808480. However, upon closer inspection
of the images from 2MASS, {\it WISE}, and the Digitized Sky Survey,
we find that the {\it WISE} source is a blend of 2MASS~J16334908$-$6808480
and a slightly fainter comoving star, 2MASS~J16334976$-$6808488.
As a result, the proper motion reported for this pair by \citet{luh14a}
is incorrect. Using the astrometry for the blend from {\it WISE} and the mean
coordinates of the pair from 2MASS, we measure ($\mu_{\alpha}$ cos $\delta$, 
$\mu_{\delta})=(-0.27\pm0.02\arcsec$~yr$^{-1}$,
$-0.32\pm0.02\arcsec$~yr$^{-1}$).
The angular separation of the pair is $3\farcs9$, corresponding to 40--50~AU
for the spectrophotometric distance estimated in Section~\ref{sec:kin}.

We have attempted to identify additional common proper motion companions
in the samples from \citet{luh14a} and Table~\ref{tab:hipm}.
For each member of those samples, we searched for stars at separations of
$\Delta\theta\leq10\arcmin$ that have a similar proper motion 
($\Delta\mu\lesssim0.05\arcsec$~yr$^{-1}$) in a catalog of previously known
high proper motion objects that was compiled for the study by \citet{luh14a}.
The resulting common proper motion pairs are listed in Table~\ref{tab:cpm},
all of which satisfy the companionship criterion of
$\Delta\theta\Delta\mu<(\mu/0.15)^{3.8}$ proposed by \citet{lep07}.

\subsection{Identifying Late-type Objects}

\begin{figure}
\epsscale{1}
\plotone{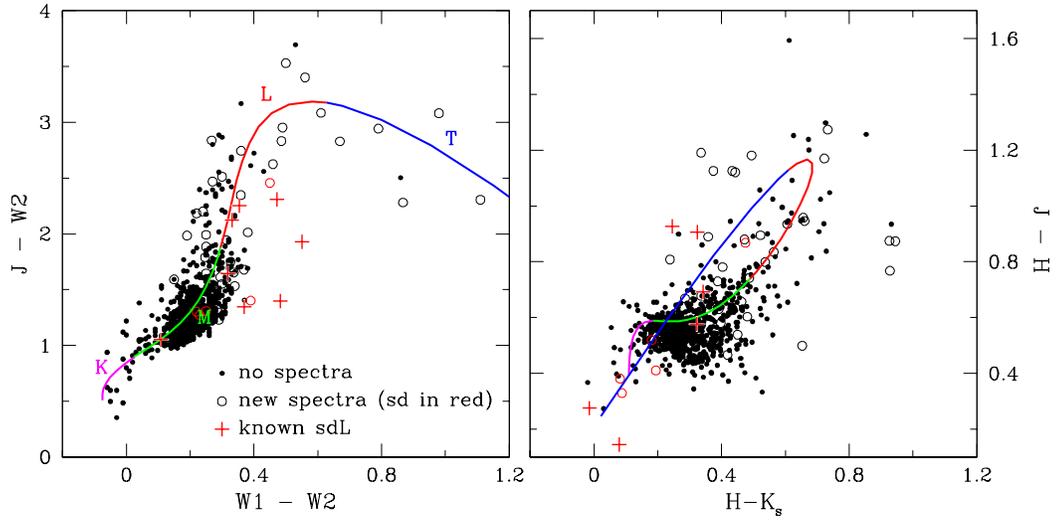}
\caption{
Color-color diagrams for high proper motion objects identified with
{\it WISE} \citep[][Table~\ref{tab:hipm}]{luh14a}, which are marked
according to whether we have obtained spectra of them (filled and open circles).
The spectroscopic targets classified as subdwarfs are plotted in red.
For comparison, we show fits to the median colors for known dwarfs
from K0--T4 (solid lines, Table~\ref{tab:fit})
and the colors of known L subdwarfs \citep[crosses,][references therein]{kir10}.
We have estimated spectral types for the high proper motion objects
based on their proximity to the fits in $J-W2$ and $W1-W2$, which
are used in Figure~\ref{fig:cmd}.
}
\label{fig:cc}
\end{figure}

The spectral types of the objects in our proper motion sample can be
estimated from their 2MASS and {\it WISE} colors. In Figure~\ref{fig:cc},
we plot $J-W2$ versus $W1-W2$ and $J-H$ versus $H-K_s$ for the sample.
The colors are indicative of spectral types ranging from K through mid-T.
To illustrate more precisely how the objects are distributed with spectral
type, we include in Figure~\ref{fig:cc} polynomial fits to the mean colors
of K dwarfs from SIMBAD and M0--T4 dwarfs from \citet{leg10a} and
\citet{kir11,kir12}.
The parameters that define these fits are presented in Table~\ref{tab:fit}.
We also show the sample of L subdwarfs compiled by \citet{kir10}, which are
significantly bluer than normal L dwarfs in $J-W2$, $J-H$, and $H-K_s$.
According to Figure~\ref{fig:cc}, most of the high proper motion objects
are probably M dwarfs. They are slightly bluer on average in $J-H$ and $H-K_s$
than the fit to normal M dwarfs, which may indicate subsolar metallicities.
The sample also appears to contain a few dozen L dwarfs and several T dwarfs.

For each object in our sample, we have identified the closest point along
the sequence of mean colors of normal dwarfs in $J-W2$ versus $W1-W2$.
The spectral types that correspond to those points are listed in
Table~\ref{tab:rpm}. 
The colors $J-H$ and $H-K_s$ were excluded from
this process because of their degeneracies with spectral type, as illustrated
in Figure~\ref{fig:cc}. Of course, these spectral types estimated
from colors should be used primarily for selecting targets of spectroscopy,
and should not be treated as true spectral types.
Because subdwarfs are bluer than normal dwarfs in $J-W2$,
their color-derived spectral types will be too early.
Nevertheless, cool subdwarfs appear at locations in a
diagram in the next section that are distinctive from those of most other
high proper motion stars, which facilitates their selection for spectroscopy.

\subsection{Identifying Nearby and Metal-poor Objects}
\label{sec:near}

We can estimate the distances of our high proper motion objects with a diagram
of $W2$ versus the color-based spectral types, as shown in Figure~\ref{fig:cmd}.
For comparison, we include a fit to absolute magnitude in $W2$, $M_{W2}$, as
a function of spectral type for M6--T4 dwarfs from \citet{dup12} and a fit
to $M_{W2}$ that we have derived for M2--M6 dwarfs from SIMBAD. These fits
are shown for distances of 10 and 20~pc. Two objects are above the fit at 10~pc.
The brighter one is WISE J104915.57$-$531906.1, which has a parallactic
distance of 2.0~pc \citep{luh13}.
The other source, WISE J163348.95$-$680851.6, appears to be an early L dwarf
at $\sim6$~pc based on Figure~\ref{fig:cmd}.
However, it is an unresolved binary in the {\it WISE} images
(Section~\ref{sec:cpm}) and the color-based spectral type is earlier than
the spectroscopic classification of M8+M8.5 (Section~\ref{sec:spec}),
resulting in an underestimate of the distance.
In Section~\ref{sec:kin}, we estimate spectrophotometric distances of
$\sim12$~pc for the components of this pair.
The color-magnitude diagram in Figure~\ref{fig:cmd} implies distances
of $\sim$10--100~pc for the remainder of the sample.

Because older stellar populations in the Galaxy have higher velocity
dispersions and tend to have lower metallicities, metal-poor stars (subdwarfs) 
typically exhibit large values of reduced proper motion, which is defined
as $H_m=m+5log(\mu)+5$ \citep{luy25} where $m$ is a particular photometric band.
Diagrams of reduced proper motion versus color or spectral type have been 
used previously to identify candidates for cool subdwarfs among high proper
motion stars \citep{kir10,kir14} and to characterize the kinematics of
known late-type dwarfs \citep{pin14}.
We have calculated $H_{W2}$ for the members of our proper motion sample.
The resulting values are presented in Table~\ref{tab:rpm} and are plotted
versus the color-based types in Figure~\ref{fig:cmd}.
We also include in Figure~\ref{fig:cmd} the known L subdwarfs compiled by
\citet{kir10} using spectral types estimated from $J-W2$ and $W1-W2$ in
the manner applied to our sample.
Because of their blue colors, the color-based spectral types for these
known subdwarfs are earlier than their true types.
One of the L subdwarfs, SDSS~J141624.08+134826.7 \citep{sch10c,bow10},
has a small value of $H_{W2}$, but the others reside near the
bottom of the reduced proper motion diagram in Figure~\ref{fig:cmd},
as expected (two are not plotted because they are below the lower limit).
Several objects in our proper motion sample also appear in that same area
of the diagram, and thus are promising candidates for cool subdwarfs.

\section{Spectroscopic Analysis}
\label{sec:spec}

\subsection{Spectroscopic Sample}

We have obtained spectra of a subset of our sample of high proper motion
objects to measure their spectral types,
focusing on those that appear to have smaller distances, cooler temperatures,
or lower metallicities based on the analysis of photometry and proper motions
in Section~\ref{sec:phot}.
We have previously presented optical spectroscopy for the primary in the
binary WISE J104915.57$-$531906.1 \citep{luh13} and near-IR spectroscopy
for both components \citep{bur13}.
That system is included among the objects marked as spectroscopic targets
in Figures~\ref{fig:cc} and \ref{fig:cmd}.
Table~\ref{tab:spec} lists the 41 additional objects that we have observed
in this study (42 if the components of WISE~J163348.95$-$680851.6 are
counted separately). Eight of our targets were also classified
spectroscopically by \citet{mac13a} and \citet{kir14}, as indicated in
Table~\ref{tab:spec}.

\subsection{SOAR/Goodman}

Six stars were observed with optical spectroscopy using the Goodman High
Throughput Spectrograph at the Southern Astrophysical Research (SOAR)
Telescope on the nights of 2013 June 2 and 2013 August 1.
The instrument was operated with the 400~l/mm grating in second order,
the GG445 filter, and the $0\farcs84$ slit, which produced a wavelength
coverage of 5400--9400~\AA\ and a resolution of 5~\AA.
The spectra were reduced using tasks within IRAF.
For the observations with SOAR and all of the other instruments in this
study, the slit was aligned near the parallactic angle for each target.

\subsection{IRTF/SpeX}

We obtained near-IR spectra of 22 objects with SpeX \citep{ray03}
at the NASA Infrared Telescope Facility (IRTF) on the nights of
January 3, June 18--20, August 25 and 26, and December 26 and 28 in 2013.
One of the SOAR targets, WISE~J194128.98$-$342335.8, was also observed again
with SpeX because no other suitable targets were available at that time.
The data were collected in the prism mode with a $0\farcs8$ slit
(0.8--2.5~\micron, $R=150$), reduced with the Spextool package \citep{cus04},
and corrected for telluric absorption \citep{vac03}. 

\subsection{Magellan/FIRE}

We performed near-IR spectroscopy on 12 targets with the 
Folded-Port Infrared Echellette \citep[FIRE;][]{sim13} at
the Magellan 6.5 m Baade Telescope at Las Campanas Observatory
on the nights of August 9 and 10 and October 27 and 28 in 2013.
We used the prism mode and the $0\farcs6$ slit. The resulting data
extended from 0.8--2.5~\micron\ and exhibited $R=500$--300 across this range.
The data were reduced with the FIREHOSE pipeline, which includes
modified versions of routines from Spextool.

\subsection{Gemini/FLAMINGOS-2}

One of our targets, WISE J210529.08$-$623558.7, was observed with FLAMINGOS-2
\citep{eik06} at the Gemini South telescope on the night of 2013 December 23.
Long-slit spectroscopy was performed with the $JH$ grism
and the $1\farcs08$ slit, which produced data from 0.9--1.8~\micron\ with
$R=200$--400. Eight 140~sec exposures were collected in an ABBA dither
pattern along the slit. The spectra were reduced with tasks in IRAF.

\subsection{SOAR/OSIRIS}

We obtained near-IR spectra of the components of WISE~J163348.95$-$680851.6 
with the Ohio State Infrared Imager/Spectrometer at SOAR on the night of
2014 January 31. We used the XD grism and the $1\arcsec$ slit, providing
data from 1.2--2.35~\micron\ with $R=1400$.  For each of the two stars,
which were well-resolved, we collected eight 30~sec exposures.
The data were reduced with IRAF.

\subsection{Spectral Classification}

Our optical spectra are presented in Figure~\ref{fig:op}. They exhibit
the strong absorption bands from TiO and VO that are indicative of M spectral
types. To classify these data, we compared them to spectra of M dwarf standards
\citep{kir91,hen94}. Each of our targets is well-matched by 
a dwarf standard. None were expected to be subdwarfs based on the
reduced proper motion diagram in Figure~\ref{fig:cmd}.
The resulting spectral types are presented in Table~\ref{tab:spec}.
The uncertainties are $\pm0.5$~subclass.

The near-IR spectra of our targets are shown in
Figures~\ref{fig:ir1}--\ref{fig:ir4}. The spectra from FLAMINGOS-2 and OSIRIS
have been smoothed to the resolution of the SpeX data.
We have omitted data at wavelengths where the correction for telluric
absorption was poor. To measure spectral types from these data,
we compared them to spectra from the SpeX Prism Spectral Libraries
for the dwarf standards adopted by \citet{kir10}.
If a good match was found, we included the spectrum of the matching standard
in Figures~\ref{fig:ir1}--\ref{fig:ir4} and adopted its type.
For the targets that did not agree with any of the dwarf standards, we
were able to find a reasonably close match within the full sample of cool
dwarfs in the Prism Libraries, which includes subdwarfs and various objects
with peculiar spectra. Those matching spectra, and the dwarf standards
with the same subclasses, are plotted with the targets in
Figures~\ref{fig:ir1}--\ref{fig:ir4}.
The spectral types derived through this process are listed in
Table~\ref{tab:spec}. They have uncertainties of $\pm1$~subclass unless
indicated otherwise.

\subsection{Comments on Individual Sources}

For the targets that lacked a good match among the dwarf standards, we 
discuss their classifications in more detail below:

{\it WISE J222409.64$-$185242.1}.
It closely resembles 2MASS J15201746$-$1755307, which has an IR spectral type
of M8 \citep{kir10}. It has deeper H$_2$O bands and a more triangular
$H$-band spectrum than the M8V standard vB~10.

{\it WISE J194128.98$-$342335.8 and WISE J235408.36+551854.5}.
They have similar spectra except that the former is slightly bluer.
They are well-matched by an average of 2MASS J15561873+1300527 and
2MASS J01151621+3130061, for which \citet{bow09} adopted d/sdM8 from
the SpeX Prism Spectral Libraries. A good match is also provided by
2MASS J18284076+1229207 \citep[M7.5 pec,][]{kir10}.

{\it WISE J001450.14$-$083823.1, WISE J204027.24+695923.7,
WISE J030601.64$-$033058.4, and WISE J043535.80+211509.2}.
The spectra of these objects are similar except that the first two
are bluer at $<$1.3~\micron\ and redder at $>$1.3~\micron.
The first two objects agree closely with 2MASS J16403197+1231068, which
has optical classifications of sdM9 \citep{giz06} and d/sdM9 \citep{bur07a}.
We adopt a type of sdM9 for all four objects, although a slightly later
type may be more appropriate for latter pair.
It is difficult to assess this possibility since late-M and early-L
subclasses for subdwarfs are sparsely sampled by available IR spectra.
\citet{kir14} has reported optical types of sdL0 for all four objects
and IR types of sdL0 for WISE J030601.64$-$033058.4 and
WISE J043535.80+211509.2.

{\it WISE J174336.62+154901.3}.
It resembles 2MASS~J14403186$-$1303263 \citep[L1 pec,][]{kir10}
except that its $K$ band is more suppressed. It is bluer than the L1
dwarf standard 2MASS J21304464$-$0845205 at $>$1.3~\micron.

{\it WISE J232219.45$-$140726.2, WISE J030845.36+325923.1,
WISE J000131.93$-$084126.9, and WISE J203751.31$-$421645.2}.
Their spectra are similar except that they become slightly bluer
from the first to the last. In addition, WISE J203751.31$-$421645.2
exhibits stronger absorption in the lines of FeH, Na~I, and K~I between
0.95--1.3~\micron\ than the others.
These objects are best matched by 2MASS J17561080+2815238
\citep[L1 pec,][]{kir10} and have deeper H$_2$O absorption and a bluer
slope at $>$1.3~\micron\ than the L1 dwarf standard 2MASS J21304464$-$0845205.

{\it WISE J195311.04$-$022954.7}.
Its spectrum is well-matched by that of 2MASS J09211410$-$2104446,
which has an optical type of L1.5 \citep{rei08} and an IR type
of L4: \citep{bur08a}. It is bluer than the L2 dwarf standard Kelu~1.

{\it WISE J103602.80+030615.6}.
It is roughly similar to 2MASS J00361617+1821104, which has
optical and IR types of L3.5 and L4, respectively \citep{kir00,bur10a}.
It is bluer at $>1.3$~\micron\ than the L4 dwarf standard
2MASS J21580457$-$1550098.

{\it WISE J000622.67$-$131955.2}.
It broadly matches the L5 dwarf standard SDSS J083506.16+195304.4, but
the $H$-band continuum exhibits a distinctive shape that may indicate
the presence of an unresolved binary. For comparison, we include
in Figure~\ref{fig:ir3} the spectrum of 2MASS J17114573+2232044,
whose spectrum was modeled as a L5+T5.5 binary \citep{bur10a}.

{\it WISE J095729.41+462413.5}.
It closely resembles 2MASS J08202996+4500315 \citep[L5,][]{kir00} and is
redder at $>1.3$~\micron\ than the L5 dwarf standard SDSS J083506.16+195304.4.

{\it WISE J214155.85$-$511853.1}.
Its H$_2$O bands and spectral slope agree well with those of
2MASS J11181292$-$0856106 \citep[L6 pec,][]{kir10}, although its lines
from FeH, Na~I, and K~I between 0.95--1.3~\micron\ are stronger. It is bluer at
$>1.3$~\micron\ than the L6 dwarf standard 2MASS J10101480$-$0406499.

{\it WISE J134310.44$-$121628.8}.
It is well-matched by 2MASS J11263991$-$5003550 \citep[L6.5 pec,][]{bur08a} and
is bluer at $>1.3$~\micron\ than the L6 dwarf standard
2MASS J10101480$-$0406499.

{\it WISE J005757.63+201304.2}.
The best available match is 2MASS J11582077+0435014 \citep[sdL7,][]{kir10}.
Because the $H$ and $K$ bands are not quite as suppressed in
WISE J005757.63+201304.2, we consider the classification of this object as
a subdwarf to be tentative based on our data. However, \citet{kir14}
do classify it as sdL7 using optical and near-IR spectra.

{\it WISE J193430.11$-$421444.3 and WISE J004713.80$-$371033.3}.
They are similar to SDSS J133148.92$-$011651.4, which has been classified
as L6, L8, and T0 \citep{haw02,kna04,sch14}. Among the dwarf standards in
that range, L9 (DENIS-P J0255$-$4700) and T0 (SDSS~J120747.17+024424.8)
provide the better matches. The $>1.3$~\micron\ slopes
of the two targets agree better with that of the T0 standard,
but they do not show the $K$-band CH$_4$ absorption of the latter.
Therefore, we adopt a spectral type of L9 for these two objects.

{\it WISE J175510.28+180320.2}.
Except for small differences in the $K$ band, it agrees well with 
SDSS J090900.73+652527.2, which has been modeled as an unresolved T1.5+T2.5
binary \citep{bur10a}. This object was originally discovered by \citet{gri12}
and \citet{mac13a}, who classified it as T2.

\section{Kinematic Analysis}
\label{sec:kin}

To examine the kinematic properties of our spectroscopic sample,
we began by estimating the distance for each object from its photometry
and the absolute magnitude expected for its spectral type.
We have done this with data in the $H$ band because $M_H$ does not differ
significantly between dwarfs and subdwarfs at a given M or L spectral type
\citep{fah12}. We adopted the values of $M_H$ produced by
the fit to $M_H$ as a function of spectral type from \citet{dup12}.
These spectrophotometric distances were then combined with the
proper motions to estimate tangential velocities.
The resulting distances and velocities are presented in Table~\ref{tab:spec}.
If an object is an unresolved binary, its distance and velocity
will be underestimated in this analysis.

Several previous studies have characterized the kinematics of samples of cool
dwarfs and subdwarfs. 
They have shown that cool dwarfs with bluer colors exhibit
higher $V_{tan}$, and hence are likely to belong to older populations with
lower metallicities \citep{fah09,sch10a}.  For instance, those measurements
have produced $V_{tan}({\rm median})\sim30$~km~s$^{-1}$
and $\sigma_{tan}\sim20$~km~s$^{-1}$ for normal L and T dwarfs \citep{fah12},
$<V_{tan}>=66$~km~s$^{-1}$ for blue L dwarfs \citep{kir10}, and
$V_{tan}\sim200$--300~km~s$^{-1}$ for M and L subdwarfs
\citep{sch09d,kir10,fah12}.
In our sample, 27 M/L dwarfs have $J-K_s$ colors that are $>0.1$~mag bluer than 
the fit to normal dwarfs from Table~\ref{tab:fit}; they exhibit
$V_{tan}({\rm median})=122$~km~s$^{-1}$. The remaining 12 M/L dwarfs
with normal or red colors have $V_{tan}({\rm median})=68$~km~s$^{-1}$.
In addition, we derive $V_{tan}({\rm median})=90$~km~s$^{-1}$ for our
11 blue L dwarfs and $V_{tan}=300$--400~km~s$^{-1}$ for our four sdM9 stars.
Thus, we find similar kinematic trends as in the previous samples.

Our estimates of tangential velocities provide a rough indication of whether
any of the objects in our sample might belong to the thick disk or halo of the
Galaxy.  Based on the simulated velocities exhibited by these populations from
\citet{dup12}, it is likely that the four sdM9 stars are members of the halo.
In addition, several objects have velocities that are suggestive of membership
in the thick disk ($V_{tan}\sim150$--200~km~s$^{-1}$), one of which is
the T1.5 dwarf WISE J210529.08$-$623558.7. It was the most promising candidate
for a thick disk/halo T dwarf prior to spectroscopy, appearing in the lower
right corner of the reduced proper motion diagram in Figure~\ref{fig:cmd}.
To examine the spectrum of WISE J210529.08$-$623558.7 for evidence of
low metallicity, we have included in Figure~\ref{fig:ir4} the model spectra
for early T dwarfs with metallicities of $[Fe/H]=0$ and $[Fe/H]=-0.5$
\citep[$T_{eff}=1400$~K, log~$g=5$,][]{burr06}. The $H$-band spectra
differ significantly between the two models while the observed
spectrum of WISE J210529.08$-$623558.7 is similar to that
of the dwarf standard. Although these data do not show an obvious signature
of low metallicity, it would be worthwhile to pursue $K$-band spectroscopy
to check for the suppression in that band that is characteristic of
cool subdwarfs. To date, only a small number of T-type objects have been found
that exhibit both kinematic and spectroscopic evidence of membership in
the thick disk or halo \citep{mac13b,pin14,burn14}.

\section{Discussion}

We have analyzed a sample of 781 high proper motion objects found with
multi-epoch astrometry from {\it WISE} \citep[Table~\ref{tab:hipm};][]{luh14a}.
Using 2MASS and {\it WISE} photometry and proper motions, we have identified
the members of this sample that are most likely to be nearby, cool, or metal
poor. We have obtained spectra of 41 of these objects, arriving at spectral
types of M4--T2.5.
Our spectroscopic sample includes 11 blue L dwarfs and five subdwarfs.
All of the latter were independently found as high proper motion objects and
classified as subdwarfs by \citet{kir14}.
Two of our candidate subdwarfs that we did not observe spectroscopically,
WISE J070720.48+170533.0 and WISE J020201.24$-$313644.7, have been confirmed
as such through spectra collected by \citet{wri14} and \citet{kir14}.
The most promising remaining candidate subdwarf from our sample
that lacks spectroscopy is WISE J141143.25$-$452418.3.

\begin{figure}
\epsscale{1}
\plotone{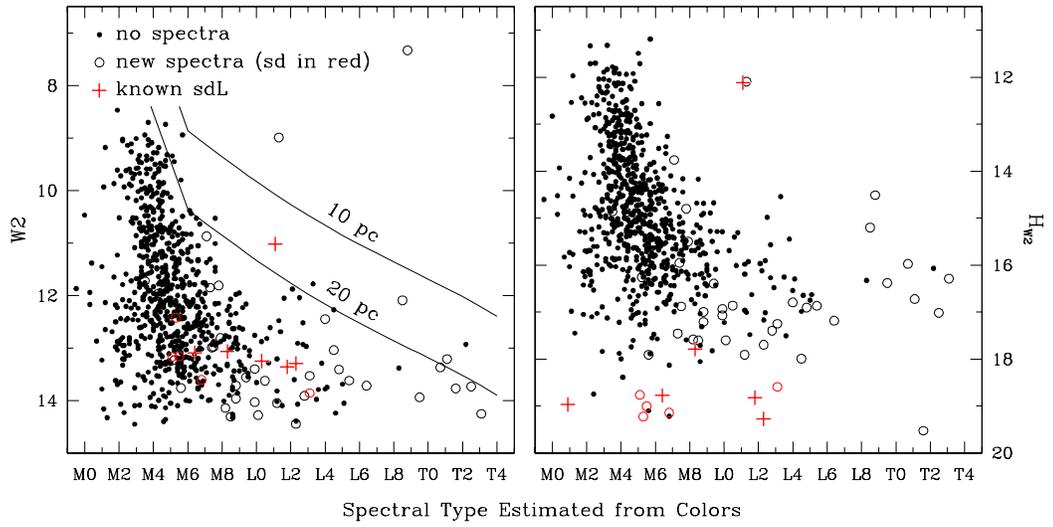}
\caption{
$W2$ and reduced proper motion in $W2$ ($H_{W2}$) versus the spectral type
estimated from $J-W2$ and $W1-W2$ (Figure~\ref{fig:cc}) for high proper
motion objects identified with {\it WISE} (filled and open circles).
In the left panel, a fit to $M_{W2}$ as a function of spectral type
is shown for 10 and 20~pc \citep[solid lines, Section~\ref{sec:near},][]{dup12}.
}
\label{fig:cmd}
\end{figure}

\begin{figure}
\epsscale{0.8}
\plotone{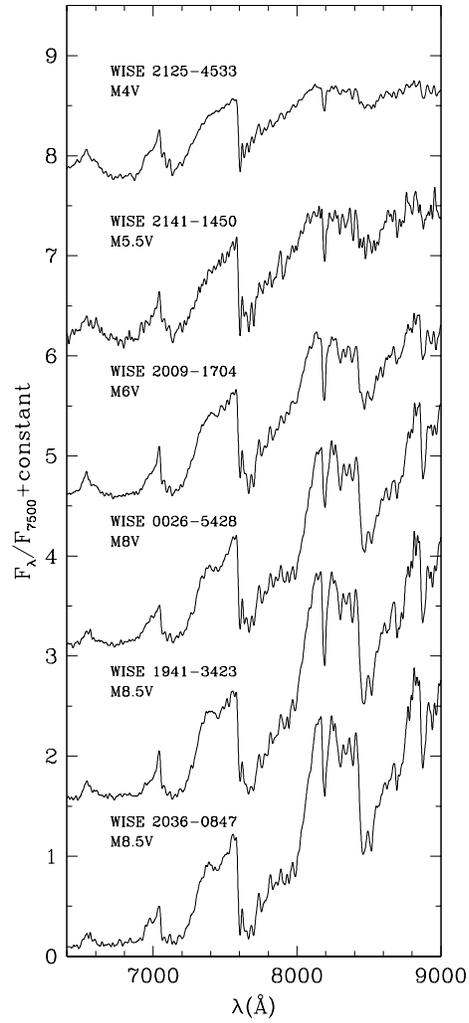}
\caption{
Optical spectra of high proper motion objects found with {\it WISE}.
The spectral types measured from these data are indicated.
The data are displayed at a resolution of 13~\AA\ and are normalized at
7500~\AA.
}
\label{fig:op}
\end{figure}

\begin{figure}
\epsscale{0.8}
\plotone{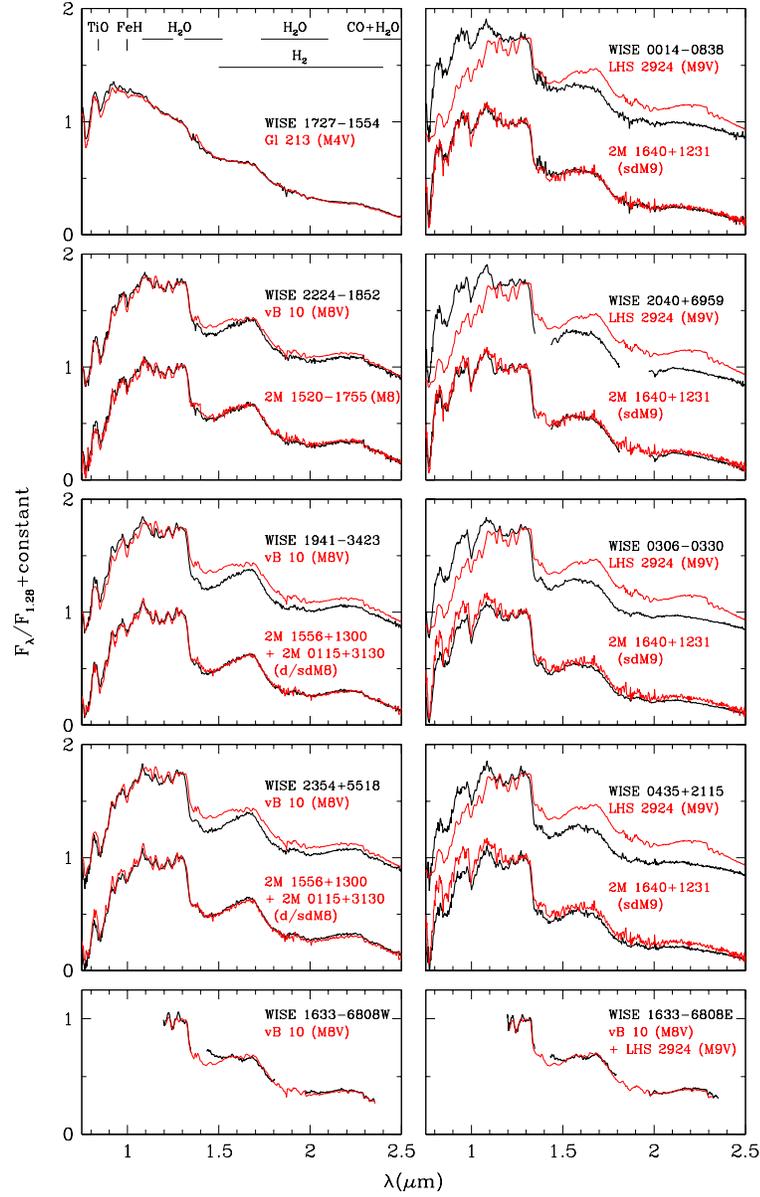}
\caption{
Near-IR spectra of high proper motion objects found with {\it WISE}
(black spectra). Each spectrum is compared to data for a dwarf standard and,
in some cases, data for another object that provide a better match
(upper and lower red spectra).
The comparison spectra are from \citet{bur04}, \citet{bur06}, \citet{kir10},
and our unpublished observations.
}
\label{fig:ir1}
\end{figure}

\begin{figure}
\epsscale{0.8}
\plotone{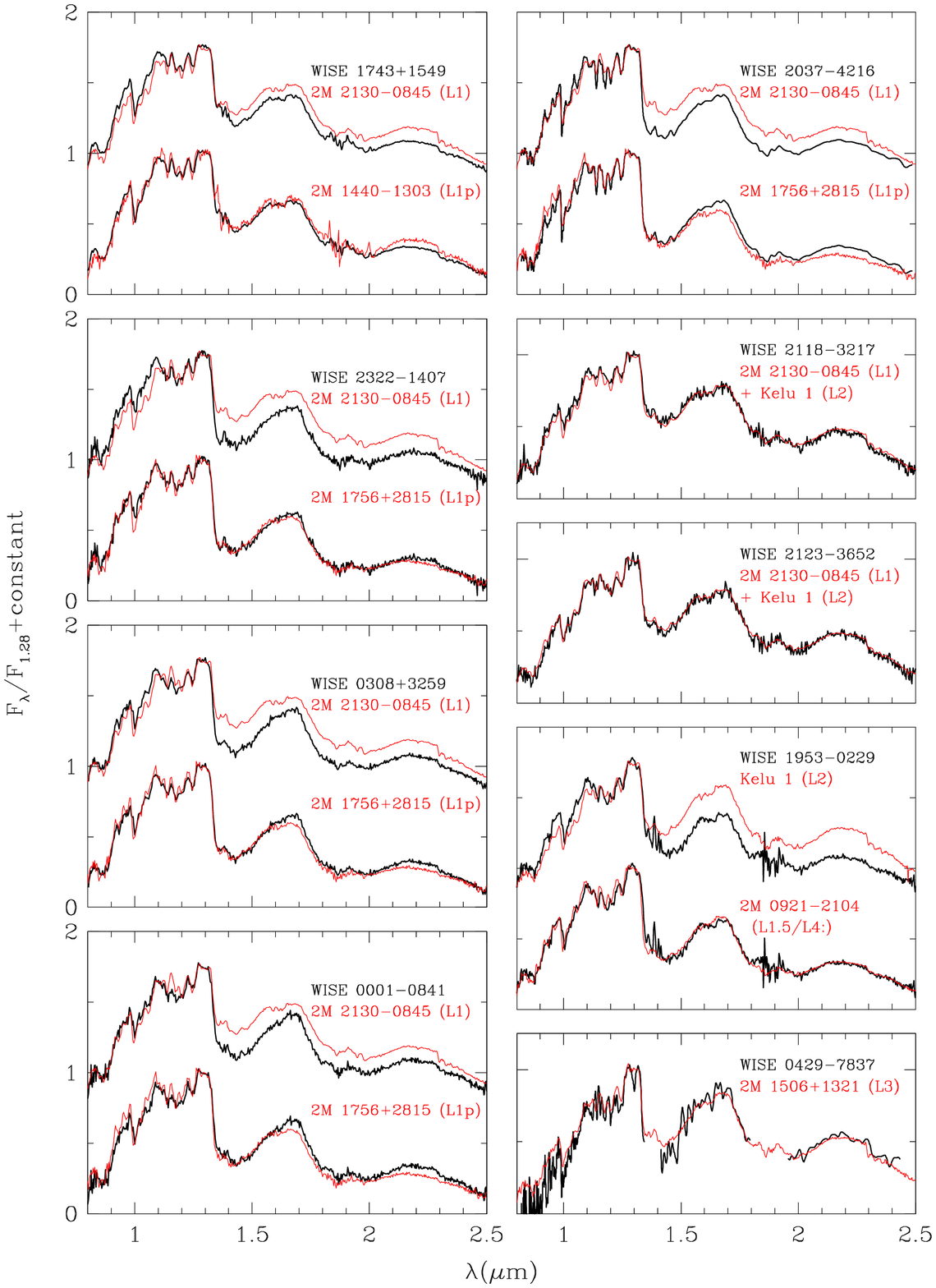}
\caption{
More near-IR spectra of high proper motion objects found with {\it WISE}
(see Fig.~\ref{fig:ir1}).
The comparison spectra are from \citet{bur07}, \citet{bur07b}, and
\citet{kir10}.
}
\label{fig:ir2}
\end{figure}

\begin{figure}
\epsscale{0.8}
\plotone{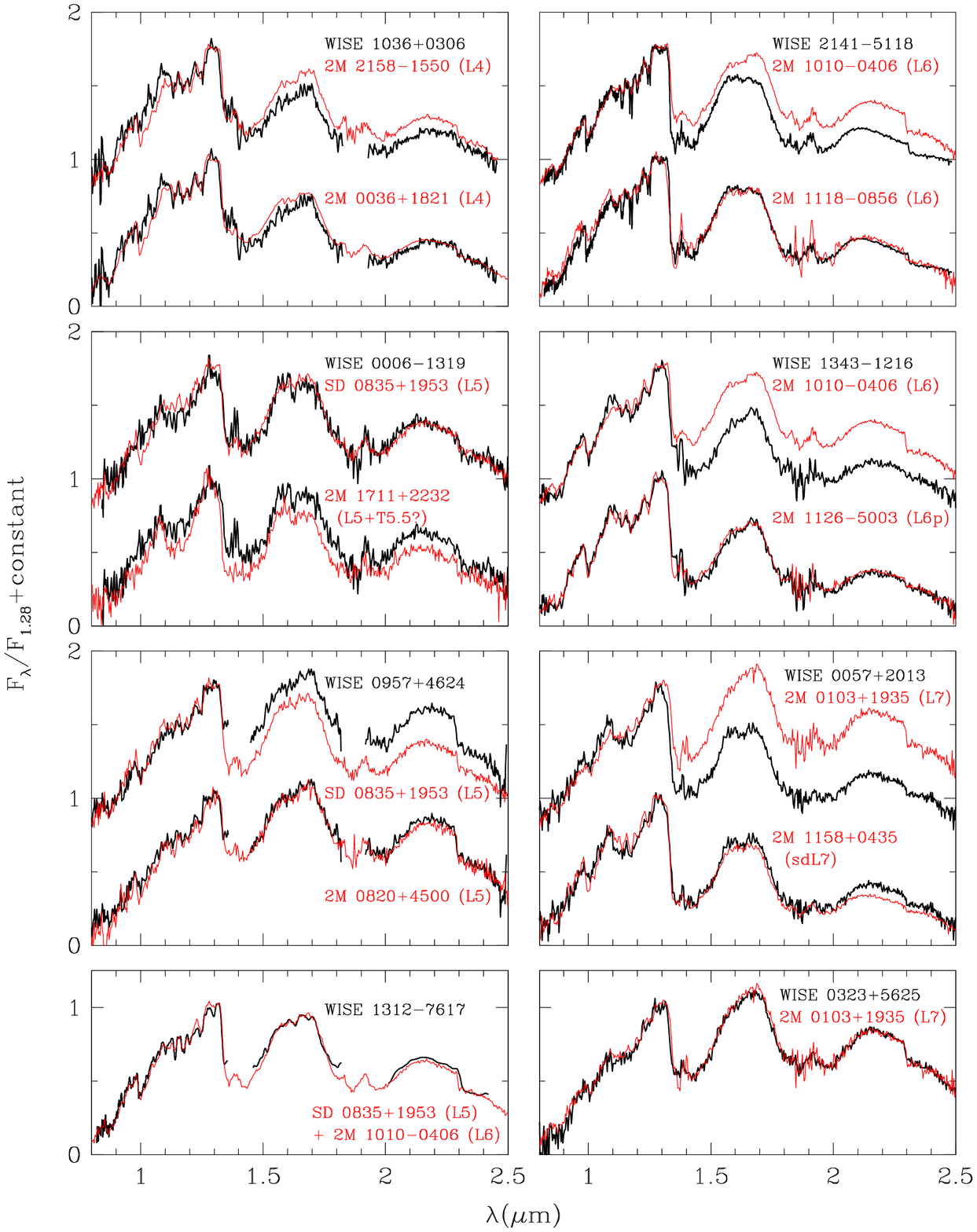}
\caption{
More near-IR spectra of high proper motion objects found with {\it WISE}
(see Fig.~\ref{fig:ir1}).
The comparison spectra are from \citet{cru04}, \citet{chi06}, \citet{rei06},
\citet{bur08b,bur10a}, and \citet{kir10}.
}
\label{fig:ir3}
\end{figure}

\begin{figure}
\epsscale{0.8}
\plotone{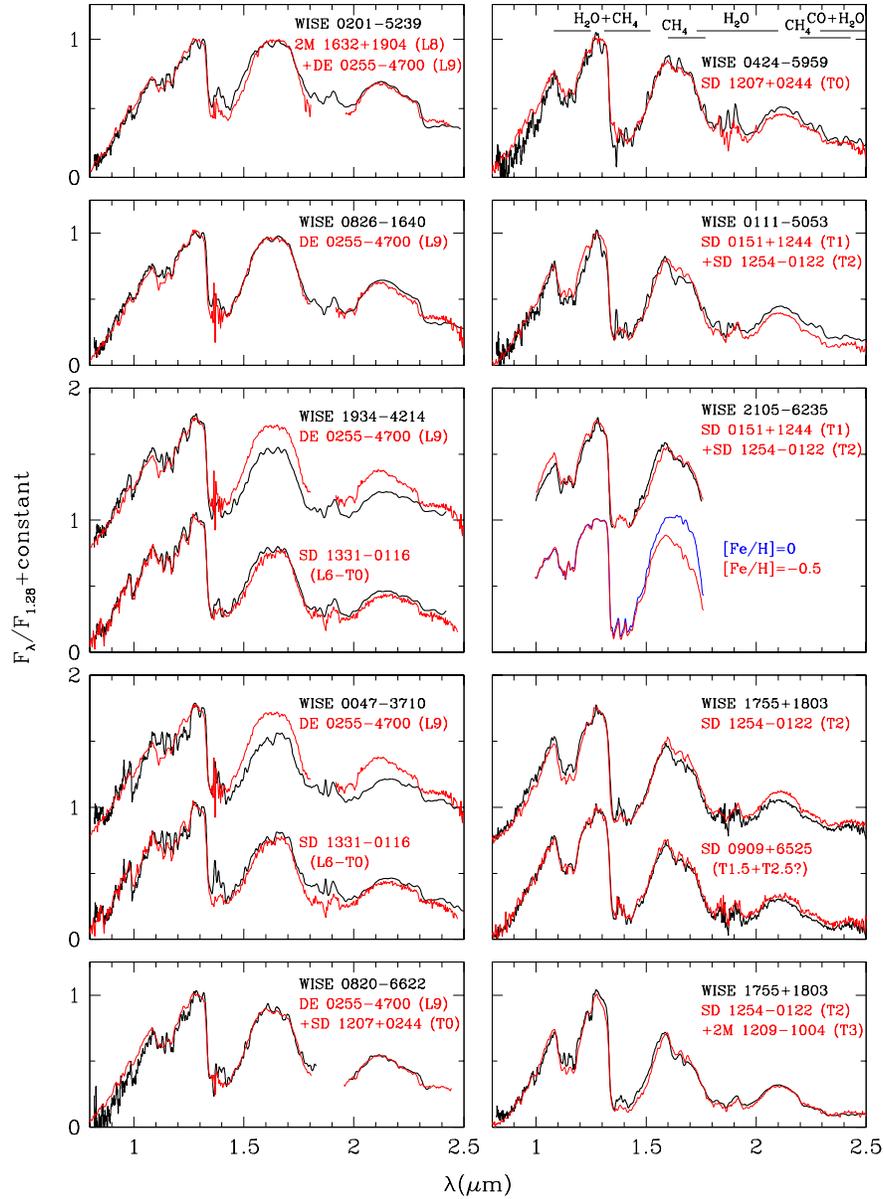}
\caption{
More near-IR spectra of high proper motion objects found with {\it WISE}
(see Fig.~\ref{fig:ir1}). The panel for WISE~2105$-$6235 includes a comparison
of model spectra from \citet{burr06} for two values of [Fe/H] at an
effective temperature of 1400~K and a surface gravity of log~$g=5$.
The comparison spectra are from \citet{bur04,bur06b,bur10a}, \citet{bur07},
\citet{chi06}, and \citet{loo07}.
}
\label{fig:ir4}
\end{figure}

We have estimated spectrophotometric distances and tangential velocities
for the members of our spectroscopic sample. The closest system appears
to have a distance of $\sim12$~pc, and several others may have distances
within 20~pc. A few objects that lack spectra also may fall within 20~pc,
as shown in Figure~\ref{fig:cmd}.
Assuming that previous samples of high proper motion stars have been
thoroughly searched for nearby stars, our proper motion survey and that
of \citet{kir14} imply that the current census of neighbors within 10~pc
has a high level of completeness for spectral types of T and earlier.
The tangential velocities in our sample are higher for bluer near-IR colors,
which is the same trend found in previous studies. Four of the five subdwarfs
exhibit velocities of 300--400~km~s$^{-1}$, which are indicative
of membership in the Galactic halo. Several additional objects have
velocities that are suggestive of the thick disk
($V_{tan}\sim150$--200~km~s$^{-1}$), including the early-T dwarf
WISE J210529.08$-$623558.7. More definitive characterizations of the kinematics
of these candidate members of the halo and thick disk
will require measurements of parallaxes and radial velocities.

\acknowledgements
K. L. acknowledges support from grant NNX12AI47G from
the NASA Astrophysics Data Analysis Program.
{\it WISE} is a joint project of the University of California, Los Angeles,
and the Jet Propulsion Laboratory (JPL)/California Institute of
Technology (Caltech), funded by NASA.
2MASS is a joint project of the University of Massachusetts and the Infrared
Processing and Analysis Center (IPAC) at Caltech, funded by NASA and the 
National Science Foundation (NSF).
The IRTF is operated by the University of Hawaii under cooperative agreement
NNX-08AE38A with NASA. The Gemini data were obtained
through program GS-2013B-DD-5. Gemini Observatory is operated by the
Association of Universities for Research in Astronomy, Inc., under a
cooperative agreement with the NSF on behalf of the Gemini partnership:
the NSF (United States), the National Research Council
(Canada), CONICYT (Chile), the Australian Research Council (Australia),
Minist\'{e}rio da Ci\^{e}ncia, Tecnologia e Inova\c{c}\~{a}o (Brazil) and
Ministerio de Ciencia, Tecnolog\'{i}a e Innovaci\'{o}n Productiva (Argentina).
The SOAR Telescope is a joint project of the Minist\'{e}rio da Ci\^{e}ncia,
Tecnologia, e Inova\c{c}\~{a}o (MCTI) da Rep\'{u}blica Federativa do Brasil,
the U.S. National Optical Astronomy Observatory (NOAO), the University of
North Carolina at Chapel Hill, and Michigan State University.
Cerro Tololo Inter-American Observatory and NOAO are operated by the
Association of Universities for Research in Astronomy, under contract with
the National Science Foundation.
This work used data from the SpeX Prism Spectral Libraries (maintained
by Adam Burgasser at http://www.browndwarfs.org/spexprism), the NASA/IPAC
Infrared Science Archive (operated by JPL under contract with NASA),
and SIMBAD database, operated at CDS, Strasbourg, France.
The Center for Exoplanets and Habitable Worlds is supported by the
Pennsylvania State University, the Eberly College of Science, and the
Pennsylvania Space Grant Consortium.

\clearpage

\begin{deluxetable}{llrrlllllll}
\tabletypesize{\scriptsize}
\tablewidth{0pt}
\tablecaption{New High Proper Motion Objects\label{tab:hipm}}
\tablehead{
\colhead{WISE} & \colhead{2MASS} & \colhead{$\mu_{\alpha}$ cos $\delta$} & \colhead{$\mu_{\delta}$} & \colhead{$J$} & \colhead{$H$} & \colhead{$K_s$} & \colhead{W1} & \colhead{W2} & \colhead{W3} & \colhead{W4}\\
\colhead{} & \colhead{} & \colhead{(arcsec~yr$^{-1}$)} & \colhead{(arcsec~yr$^{-1}$)} & \colhead{(mag)} & \colhead{(mag)} & \colhead{(mag)} & \colhead{(mag)} & \colhead{(mag)} & \colhead{(mag)} & \colhead{(mag)}}
\startdata
J000131.93$-$084126.9 & J00013166$-$0841234 & 0.331$\pm$0.014 & $-$0.299$\pm$0.014 & 15.71$\pm$0.05 & 15.03$\pm$0.06 & 14.70$\pm$0.09 & 14.29$\pm$0.03 & 13.96$\pm$0.05 & \nodata & \nodata \\
J020110.68$-$523916.9 & J02011020$-$5239186 & 0.418$\pm$0.014 & 0.155$\pm$0.015 & 16.45$\pm$0.11 & 15.58$\pm$0.10 & 14.63$\pm$0.10 & 14.11$\pm$0.03 & 13.62$\pm$0.04 & \nodata & \nodata \\
J030845.36+325923.1 & J03084507+3259277 & 0.310$\pm$0.011 & $-$0.376$\pm$0.012 & 15.80$\pm$0.06 & 15.19$\pm$0.07 & 14.71$\pm$0.07 & 14.48$\pm$0.03 & 14.14$\pm$0.06 & \nodata & \nodata \\
J045425.37+400408.5 & J04542499+4004106 & 0.386$\pm$0.011 & $-$0.169$\pm$0.012 & 15.38$\pm$0.05 & 14.97$\pm$0.08 & 14.50$\pm$0.08 & 14.37$\pm$0.04 & 14.33$\pm$0.09 & \nodata & \nodata \\
J060609.90$-$145318.3 & \nodata & 0.136$\pm$0.010 & $-$0.200$\pm$0.010 & 12.80$\pm$0.03 & 12.31$\pm$0.02 & 12.07$\pm$0.02 & 11.93$\pm$0.02 & 11.79$\pm$0.02 & \nodata & \nodata \\
J061700.64+040050.0 & J06170068+0400548 & $-$0.057$\pm$0.010 & $-$0.459$\pm$0.010 & 14.29$\pm$0.03 & 13.79$\pm$0.04 & 13.39$\pm$0.03 & 13.27$\pm$0.03 & 12.99$\pm$0.03 & \nodata & \nodata \\
J080457.04$-$374622.1 & J08045688$-$3746183 & 0.200$\pm$0.012 & $-$0.331$\pm$0.010 & 14.69$\pm$0.03 & 14.15$\pm$0.04 & 13.73$\pm$0.05 & 13.47$\pm$0.03 & 13.18$\pm$0.03 & \nodata & \nodata \\
J095729.41+462413.5 & J09572983+4624177 & $-$0.354$\pm$0.011 & $-$0.353$\pm$0.011 & 16.25$\pm$0.11 & 15.37$\pm$0.13 & 14.45$\pm$0.09 & 13.68$\pm$0.03 & 13.41$\pm$0.04 & \nodata & \nodata \\
J103602.80+030615.6 & J10360307+0306160 & $-$0.403$\pm$0.015 & $-$0.043$\pm$0.014 & 15.97$\pm$0.10 & 15.48$\pm$0.11 & 14.82$\pm$0.12 & 14.35$\pm$0.03 & 14.03$\pm$0.05 & \nodata & \nodata \\
J145209.23$-$423545.6 & \nodata & $-$0.260$\pm$0.010 & $-$0.153$\pm$0.010 & 13.09$\pm$0.10\tablenotemark{a} & 12.45$\pm$0.10\tablenotemark{a} & 12.34$\pm$0.10\tablenotemark{a} & 12.23$\pm$0.03 & 12.13$\pm$0.03 & \nodata & \nodata \\
J151612.81$-$550826.5 & \nodata & $-$0.227$\pm$0.010 & $-$0.144$\pm$0.010 & 13.99$\pm$0.03 & 13.45$\pm$0.03 & 13.30$\pm$0.04 & 13.23$\pm$0.03 & 13.22$\pm$0.03 & \nodata & \nodata \\
J175510.28+180320.2 & J17551062+1803203 & $-$0.453$\pm$0.014 & $-$0.008$\pm$0.014 & 16.02$\pm$0.09 & 15.22$\pm$0.09 & 14.68$\pm$0.13 & 14.60$\pm$0.03 & 13.73$\pm$0.04 & \nodata & \nodata \\
J184936.22$-$204538.1 & J18493649$-$2045352 & $-$0.313$\pm$0.010 & $-$0.247$\pm$0.010 & 13.83$\pm$0.03 & 13.30$\pm$0.03 & 13.05$\pm$0.04 & 12.97$\pm$0.03 & 12.80$\pm$0.04 & \nodata & \nodata \\
J193430.11$-$421444.3 & J19343030$-$4214401 & $-$0.204$\pm$0.024 & $-$0.397$\pm$0.025 & 16.79$\pm$0.17 & 15.67$\pm$0.16 & 15.29$\pm$0.17 & 14.81$\pm$0.04 & 14.45$\pm$0.07 & \nodata & \nodata \\
J195311.04$-$022954.7 & J19531118$-$0229501 & $-$0.187$\pm$0.010 & $-$0.403$\pm$0.012 & 15.64$\pm$0.05 & 14.86$\pm$0.07 & 14.45$\pm$0.07 & 14.01$\pm$0.03 & 13.62$\pm$0.04 & \nodata & \nodata \\
J203644.55$-$084715.1 & J20364425$-$0847138 & 0.379$\pm$0.010 & $-$0.118$\pm$0.010 & 13.42$\pm$0.03 & 12.88$\pm$0.03 & 12.43$\pm$0.02 & 12.10$\pm$0.02 & 11.81$\pm$0.02 & 11.43$\pm$0.19 & \nodata \\
J203751.31$-$421645.2 & J20375108$-$4216410 & 0.229$\pm$0.010 & $-$0.391$\pm$0.010 & 15.50$\pm$0.05 & 14.76$\pm$0.06 & 14.27$\pm$0.06 & 13.96$\pm$0.03 & 13.71$\pm$0.04 & \nodata & \nodata \\
J211157.85$-$521111.2 & \nodata & $-$0.237$\pm$0.031 & 0.095$\pm$0.032 & 16.56$\pm$0.17 & 15.92$\pm$0.21 & \nodata & 15.36$\pm$0.05 & 14.25$\pm$0.05 & \nodata & \nodata \\
J214152.86$-$145013.4 & J21415277$-$1450092 & 0.115$\pm$0.011 & $-$0.386$\pm$0.010 & 14.46$\pm$0.04 & 13.97$\pm$0.04 & 13.71$\pm$0.05 & 13.52$\pm$0.03 & 13.23$\pm$0.04 & \nodata & \nodata \\
J232219.45$-$140726.2 & J23221914$-$1407238 & 0.399$\pm$0.020 & $-$0.221$\pm$0.019 & 15.99$\pm$0.07 & 15.52$\pm$0.11 & 15.10$\pm$0.14 & 14.68$\pm$0.04 & 14.31$\pm$0.06 & \nodata & \nodata \\
\enddata
\tablenotetext{a}{The high proper motion star is blended with an artifact in
the 2MASS images, but it appears to be the dominant source of flux.}
\tablecomments{$J$, $H$, and $K_s$ are from the 2MASS Survey Point Source Reject
Table for WISE~J060609.90$-$145318.3, WISEA~J151612.81$-$550826.5, and
WISE~J211157.85$-$521111.2, from our manual measurement of photometry from
the 2MASS images for WISE~J145209.23$-$423545.6, and from the 2MASS Point
Source Catalog for all other sources.
W1--W4 are from the AllWISE Source Catalog for WISEA~J151612.81$-$550826.5 
and from {\it WISE} All-Sky Source Catalog for all other sources. {\it WISE}
detections that appear false or unreliable based on visual inspection have
been omitted. Proper motions are based on astrometry from 2MASS and {\it WISE}.
The 2MASS astrometry for WISE~J145209.23$-$423545.6 was measured manually from
those images.}
\end{deluxetable}

\begin{deluxetable}{lll}
\tabletypesize{\scriptsize}
\tablewidth{0pt}
\tablecaption{Common Proper Motion Pairs\label{tab:cpm}}
\tablehead{
\colhead{New High Proper} & \colhead{Known Star with Similar} & \colhead{Separation}\\
\colhead{Motion Object} & \colhead{Proper Motion} & \colhead{(arcsec)}}
\startdata
WISE J035227.54$-$315104.9\tablenotemark{a} & LHS 1609 & 40 \\
WISE J051723.87$-$345121.8\tablenotemark{a} & HD 34642 & 159 \\
WISE J064837.92+073658.4 & HD 49409 & 28 \\
WISE J065717.78$-$144641.2 & LP 721-15 & 43 \\
WISE J111614.19$-$440325.2 & LHS 2386 & 355 \\
WISE J124014.80+204752.8\tablenotemark{a} & BD+21 2442 & 113 \\
WISE J141420.38$-$055709.1 & HD 124553 & 21 \\
WISE J145824.02$-$390754.7 & SCR J1457$-$3904 & 446 \\
WISE J154803.74$-$581055.5 & LHS 3119 & 21 \\
WISE J190756.42$-$151416.3 & HD 178140 & 22 \\
WISE J191046.06$-$413348.7 & 2MASS J19103460$-$4133443 & 128 \\
WISE J200436.76$-$712356.6 & LTT 7914 & 39 \\
WISE J202218.29$-$582110.3 & WD 2018$-$585 & 65 \\
WISE J222742.02$-$233733.4 & LP 876-1 & 64 \\
\enddata
\tablenotetext{a}{The common proper motion of this pair was previously noted
by \citet{kir14}.}
\end{deluxetable}

\begin{deluxetable}{lllllll}
\tabletypesize{\scriptsize}
\tablewidth{0pt}
\tablecaption{Coefficients of Polynomial Fits to Colors for Dwarfs from K0--T4\label{tab:fit}}
\tablehead{
\colhead{$y$} & \colhead{$c_0$} & \colhead{$c_1$} & \colhead{$c_2$} & \colhead{$c_3$} & \colhead{$c_4$} & \colhead{$c_5$}}
\startdata
$J-H$ & 0.5879725 & $-4.185039\times10^{-4}$ & $-8.971617\times10^{-4}$ &  $3.156630\times10^{-4}$ &  $1.088121\times10^{-6}$ & $-5.701197\times10^{-7}$ \\
$H-K_s$ & 0.1943348 &   $1.967872\times10^{-2}$ &  $1.172085\times10^{-3}$ &  $1.991087\times10^{-5}$ &  $1.084605\times10^{-7}$ & $-2.049361\times10^{-7}$ \\
$J-W2$ & 0.9256119  &  $5.609014\times10^{-2}$ &  $2.061348\times10^{-3}$ &  $3.043627\times10^{-4}$ &  $5.493171\times10^{-5}$ & $-9.495056\times10^{-7}$ \\
$W1-W2$ & $4.016049\times10^{-2}$ &   $2.857693\times10^{-2}$ &  $9.994479\times10^{-4}$ & $-1.337646\times10^{-4}$ & $-1.948121\times10^{-6}$ &  $3.257249\times10^{-7}$  \\
\enddata
\tablecomments{$y=\Sigma_{i=0} c_{i}x^i$ where $x=-8$, 0, 10, and 20
for K0, M0, L0, and T0, respectively.}
\end{deluxetable}

\begin{deluxetable}{llc}
\tablewidth{0pt}
\tablecaption{Reduced Proper Motions and Spectral Type Estimates\label{tab:rpm}}
\tablehead{
\colhead{WISE} & \colhead{$H_{W2}$} & \colhead{Spectral Type} \\
\colhead{} & \colhead{} & \colhead{Estimated from Colors\tablenotemark{a}}}
\startdata
J000131.93$-$084126.9 & 17.21 & 8.8 \\
J000622.67$-$131955.2 & 17.18 & 16.4 \\
J000721.39+500302.7 & 16.54 & 4.7 \\
J001318.52$-$554802.3 & 16.79 & 6.5 \\
J001450.14$-$083823.1 & 19.01 & 5.5 \\
\enddata
\tablecomments{
This table is available in its entirety in a machine-readable form in the
online journal. A portion is shown here for guidance regarding its form and
content.}
\tablenotetext{a}{Numerical types are defined such that $-8$, 0, 10, and 20
correspond to K0, M0, L0, and T0, respectively.}
\end{deluxetable}

\begin{deluxetable}{llllll}
\tablewidth{0pt}
\tablecaption{Spectral Types, Distances, and Tangential Velocities of High Proper Motion Objects\label{tab:spec}}
\tabletypesize{\scriptsize}
\tablehead{
\colhead{WISE} &
\colhead{Telescope/Instrument} &
\colhead{Date} &
\colhead{Spectral} &
\colhead{Distance\tablenotemark{a}} &
\colhead{$V_{tan}$\tablenotemark{b}}\\
\colhead{} &
\colhead{} &
\colhead{} &
\colhead{Type} &
\colhead{(pc)} &
\colhead{(km/s)}}
\startdata
J000131.93$-$084126.9 & IRTF/SpeX & 2013 Aug 26 & L1 pec (blue) & 61$\pm$13 & 129$\pm$29 \\
J000622.67$-$131955.2 & IRTF/SpeX & 2013 Jan 3 & L5 pec & 41$\pm$9 & 95$\pm$21 \\
J001450.14$-$083823.1 & IRTF/SpeX & 2013 Dec 28 & sdM9\tablenotemark{c} & 47$\pm$10 & 328$\pm$69 \\
J002656.73$-$542854.8 & SOAR/Goodman & 2013 Aug 1 & M8V & 19$\pm$4 & 34$\pm$6 \\
J004713.80$-$371033.3 & Magellan/FIRE & 2013 Oct 27 & L9 pec (blue)\tablenotemark{d} & 23$\pm$4 & 64$\pm$11 \\
J005757.63+201304.2 & IRTF/SpeX & 2013 Dec 28 & sdL7?\tablenotemark{e} & 29$\pm$6 & 122$\pm$24 \\
J011154.36$-$505343.2 & Magellan/FIRE & 2013 Oct 27 & T1.5 & 17$\pm$2 & 40$\pm$5 \\
J020110.68$-$523916.9 & Magellan/FIRE & 2013 Aug 9 & L8.5 & 26$\pm$5 & 56$\pm$10 \\
J030601.64$-$033058.4 & IRTF/SpeX & 2013 Dec 26 & sdM9\tablenotemark{c} & 49$\pm$10 & 307$\pm$65 \\
J030845.36+325923.1 & IRTF/SpeX & 2013 Aug 26 & L1 pec (blue) & 65$\pm$15 & 151$\pm$34 \\
J032301.86+562558.0 & IRTF/SpeX & 2013 Dec 26 & L7 & 18$\pm$3 & 35$\pm$7 \\
J042449.21$-$595905.6 & Magellan/FIRE & 2013 Oct 28 & T0 & 28$\pm$4 & 40$\pm$6 \\
J042949.41$-$783705.6 & Magellan/FIRE & 2013 Oct 28 & L3$\pm$2 & 54$\pm$19 & 128$\pm$45 \\
J043535.80+211509.2 & IRTF/SpeX & 2013 Dec 26 & sdM9\tablenotemark{c} & 66$\pm$14 & 398$\pm$84 \\
J082000.48$-$662211.9 & Magellan/FIRE & 2013 Oct 28 & L9.5 & 21$\pm$3 & 33$\pm$4 \\
J082640.46$-$164032.0 & Magellan/FIRE & 2013 Oct 27 & L9\tablenotemark{f} & 16$\pm$3 & 75$\pm$13 \\
J095729.41+462413.5 & IRTF/SpeX & 2013 Jun 18 & L5 pec (red) & 38$\pm$9 & 89$\pm$20 \\
J103602.80+030615.6 & IRTF/SpeX & 2013 Jun 19 & L4 pec (blue) & 47$\pm$11 & 90$\pm$20 \\
J131211.10$-$761740.9 & Magellan/FIRE & 2013 Aug 10 & L5.5 & 33$\pm$7 & 87$\pm$19 \\
J134310.44$-$121628.8 & IRTF/SpeX & 2013 Jun 19 & L6.5$\pm$2 pec (blue) & 31$\pm$10 & 68$\pm$21 \\
J163348.95$-$680851.6 & SOAR/OSIRIS & 2014 Jan 31 & M8V+M8.5V\tablenotemark{g} & 12$\pm$2 & 23$\pm$5 \\
J172733.09$-$155414.2 & IRTF/SpeX & 2013 Jun 20 & M4V & 64$\pm$31 & 128$\pm$63 \\
J174336.62+154901.3 & IRTF/SpeX & 2013 Jun 20 & L1 pec (blue) & 34$\pm$8 & 56$\pm$12 \\
J175510.28+180320.2 & IRTF/SpeX & 2013 Jun 18 & T2\tablenotemark{h} & 19$\pm$2 & 40$\pm$5 \\
J193430.11$-$421444.3 & Magellan/FIRE & 2013 Aug 9 & L9 pec (blue) & 26$\pm$5 & 56$\pm$10 \\
J194128.98$-$342335.8 & SOAR/Goodman & 2013 Aug 1 & M8.5V & 50$\pm$9 & 143$\pm$26 \\
 & IRTF/SpeX & 2013 Aug 25 & M8 pec (blue) & 53$\pm$12 & 152$\pm$34 \\
J195311.04$-$022954.7 & IRTF/SpeX & 2013 Jun 19 & L2 pec (blue) & 49$\pm$11 & 102$\pm$24 \\
J200907.88$-$170448.0 & SOAR/Goodman & 2013 Aug 1 & M6V & 103$\pm$24 & 330$\pm$78 \\
J203644.55$-$084715.1 & SOAR/Goodman & 2013 Aug 1 & M8.5V & 30$\pm$6 & 57$\pm$11 \\
J203751.31$-$421645.2 & Magellan/FIRE & 2013 Aug 9 & L1 pec (blue) & 54$\pm$12 & 115$\pm$26 \\
J204027.24+695923.7 & IRTF/SpeX & 2013 Dec 28 & sdM9\tablenotemark{c} & 35$\pm$7 & 380$\pm$81 \\
J210529.08$-$623558.7 & Gemini/FLAMINGOS-2 & 2013 Dec 23 & T1.5 & 26$\pm$4 & 176$\pm$25 \\
J211157.85$-$521111.2 & Magellan/FIRE & 2013 Aug 9 & T2.5 & 25$\pm$4 & 31$\pm$5 \\
J211807.07$-$321713.5 & IRTF/SpeX & 2013 Aug 25 & L1.5 & 46$\pm$10 & 80$\pm$18 \\
J212354.78$-$365223.4 & IRTF/SpeX & 2013 Aug 25 & L1.5 & 45$\pm$10 & 109$\pm$25 \\
J212502.66$-$453353.5\tablenotemark{i} & SOAR/Goodman & 2013 Aug 1 & M4V & 73$\pm$21 & 132$\pm$39 \\
J214152.86$-$145013.4 & SOAR/Goodman & 2013 Aug 1 & M5.5V & 136$\pm$40 & 261$\pm$77 \\
J214155.85$-$511853.1 & Magellan/FIRE & 2013 Aug 9 & L6 pec (blue) & 19$\pm$4 & 65$\pm$14 \\
J222409.64$-$185242.1 & IRTF/SpeX & 2013 Aug 25 & M8 & 52$\pm$12 & 96$\pm$21 \\
J232219.45$-$140726.2 & IRTF/SpeX & 2013 Aug 25 & L1 pec (blue) & 76$\pm$17 & 165$\pm$37 \\
J235408.36+551854.5 & IRTF/SpeX & 2013 Dec 26 & M8 pec (blue) & 31$\pm$7 & 196$\pm$44 \\
\enddata
\tablenotetext{a}{Spectrophotometric distances based on the 2MASS $H$-band
photometry and the value of $M_H$ produced by the relation between spectral
type and $M_H$ from \citet{dup12} except for the two stars classified
as M4V, for which the median $M_H$ for M4 dwarfs within 8~pc was used
\citep{kir12}.
}
\tablenotetext{b}{Estimated from the spectrophotometric distances
and the proper motions in Table~\ref{tab:hipm} and \citet{luh14a}.}
\tablenotetext{c}{Classified as sdL0 by \citet{kir14}.}
\tablenotetext{d}{Classified as L4: by \citet{kir14}.}
\tablenotetext{e}{Classified as sdL7 by \citet{kir14}.}
\tablenotetext{f}{Classified as L9 by \citet{kir14}.}
\tablenotetext{g}{The spectral types apply to the west and east components,
respectively.}
\tablenotetext{h}{Classified as T2 by \citet{mac13a}.}
\tablenotetext{i}{Originally identified as a high proper motion star by
\citet{pok04}.}
\end{deluxetable}

\clearpage

\end{document}